\documentclass[12pt]{spieman}  
\usepackage{amsmath,amsfonts,amssymb}
\usepackage{graphicx}
\usepackage{setspace}
\usepackage{tocloft}

\title{Pharmacokinetic parameters quantification in DCE-MRI for prostate cancer}

\author[a]{Jhonalbert Aponte}
\author[a]{Álvaro Ruiz}
\author[b]{Jacksson Sánchez}
\author[c,d,*]{Miguel Martín-Landrove}
\affil[a]{Servicio de Radioterapia Avanzada, Fundación Arturo López Pérez, Santiago de Chile, Chile}
\affil[b]{Faculty of Science and Technology, Physics Department, Universidad Nacional Pedro Henríquez Ureña, Santo Domingo, Dominican Republic}
\affil[c]{Centre for Medical Visualization, National Institute for Bioengineering, Universidad Central de Venezuela, Caracas, Venezuela}
\affil[d]{Centro de Diagnóstico Docente Las Mercedes, Caracas, Venezuela}

\cftpagenumbersoff{figure}
\cftpagenumbersoff{table} 
\begin{document} 
\maketitle

\begin{abstract}
Tumor vascularity detection and quantification are of high relevance in the assessment of cancer lesions not only for disease diagnostics but for therapy considerations and monitoring. The present work addressed the quantification of pharmacokinetic parameters derived from the two-compartment Brix model by analyzing and processing Dynamic Contrast-Enhanced Magnetic Resonance Images (DCE-MRI) of prostate cancer lesions. The 3D image sets were acquired at regular time intervals, covering all the phases implied in contrast injection (wash-in and wash-out phases), and the standardized image intensity is determined for each voxel, conforming to a 4D data set. Previous voxel classification was carried out by the three-time-point method proposed by Degani et al. (1997) and Furman-Haran et al. (1998) to identify regions of interest. Relevant pharmacokinetic parameters, such as $k_{el}$, the vascular elimination rate, and $k_{ep}$, the extravascular transfer rate, are extracted by a novel interpolation method applicable to compartment models. Parameter distribution maps were obtained for either pathological or unaffected glandular regions indicating that a three-compartment model, including fast and slow exchange compartments, provides a more suitable description of the contrast kinetics. Results can be applied to prostate cancer diagnostic evaluation and therapy follow-up. 
\end{abstract}


{\noindent \footnotesize\textbf{*}Miguel Martín-Landrove,  \linkable{mglmrtn@gmail.com} }

\begin{spacing}{2}   

\section{Introduction}
\label{sect:intro}  
Dynamic Contrast-Enhanced MRI or DCE-MRI has been applied extensively to diagnose and quantify several pathologies associated with cancer \cite{Choyke03,Alonzi07,Haris08,Foottit10,Verma12,Chen14}. It essentially consists of the controlled intravenous delivery of a contrast agent, typically a Gadolinium compound, followed by the acquisition of a volumetric $T_{1}$-weighted MRI. The resulting image dataset can be analyzed either qualitatively \cite{Degani97,Hauth06,Lenkinski11,Fluckiger12,Lavini13,Furman14} or quantitatively \cite{Foottit10,Verma12,Chen14,Sourbron09,Koh11}. Several compartmental models are proposed for the pharmacokinetics of the contrast agent \cite{Koh11,Brix91,Tofts91,Tofts97,Fusco12}. In the case of prostate DCE-MRI, pharmacokinetic two-compartment models are applicable \cite{Verma12,Brix91,Tofts91}. The present work is addressed to the quantification of pharmacokinetic parameters, as derived from the two-compartment Brix-Tofts model \cite{Brix91,Tofts91} and multi-compartmental models \cite{Ambi16,Chen11,Chen111,Wang06}, through the analysis and processing of DCE-MRI of prostate cancer lesions, using the Levenberg-Marquardt algorithm \cite{Leven44,Marq63} and modified de Prony method \cite{dProny1795,Mml07}.

The work is organized as follows, in Section 2, general aspects of the two-compartment Brix-Tofts model, three-compartment models, image acquisition and qualitative analysis of the image data set, and quantitative evaluation of the pharmacokinetic parameters are discussed; in Section 3, the results are discussed and some conclusions are presented. 

\section{Materials and methods}

\subsection{Two-compartment Brix-Tofts model}
\begin{figure}[!h]
    \centering
    \includegraphics[width=10cm]{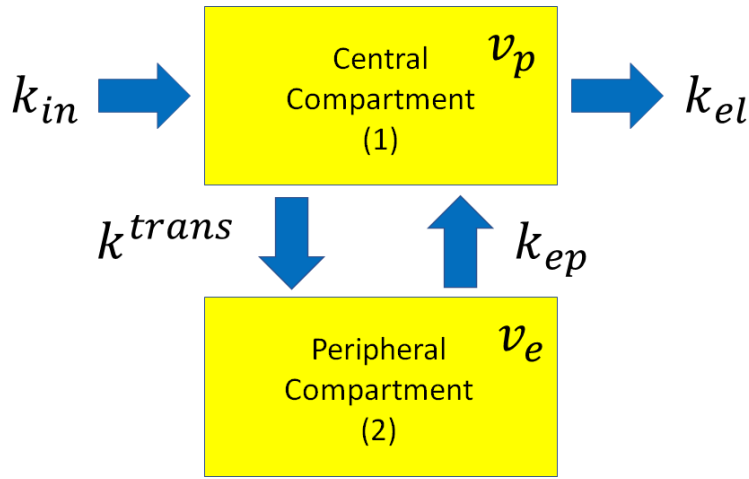}
    \caption{Schematic representation of e two-compartment Brix-Tofts model}
    \label{fig:f1}
\end{figure}

Essentially, the model establishes that the contrast intake occurs in two connected environments or compartments: one, which is called the central compartment related to the capillary and vascular space, and the second, the peripheral compartment, generally related to extravascular and extracellular space. The model can be depicted schematically as shown in Figure \ref{fig:f1}, after \cite{Brix91},

The parameters of the model $k_{in}$, $k^{trnas}$, $k_{ep}$, $k_{el}$, $v_p$ and $v_e$ describe the pharmacokinetics in the system, i.e., concentration exchange velocities and volumes of the compartments, respectively. The differential equations that describe the pharmacokinetics are,
\begin{eqnarray}\label{eq:eq1}
\nonumber
	&\frac{dM_1}{dt}=&k_{in}-(k^{trans}+k_{el})M_1+k_{ep}M_2
	\\*
	&\frac{dM_2}{dt}=&k^{trans}M_1-k_{ep}M_2
\end{eqnarray}
where $M_1$ and $M_2$ are the contrast agent total mass in compartments 1 and 2 respectively. Assuming that the transfer velocity between both compartments is equal,
\begin{equation}
	k^{trans}v_p=k_{ep}v_e
\end{equation}
and $v_e \ll v_p$, equation \ref{eq:eq1} can be written,
\begin{equation}\label{eq:eq3}
	\frac{dC_1}{dt}=\frac{k_{in}}{v_p}-k_{el}C_1,
	\\
	\frac{dC_2}{dt}=\frac{v_p}{v_e}k^{trans}C_1-k_{ep}C_2
\end{equation}
where $C_1$ and $C_2$ are the contrast concentrations in the compartments. Because the contrast agent is delivered as a bolus, the set defined in equation \ref{eq:eq3} has a solution,
\begin{eqnarray}\label{eq:eq4}
\nonumber
	&C_1(t)=&\frac{k_{in}}{v_pk_{el}}(e^{k_{el}t'}-1)e^{-k_{el}t}
	\\*
\nonumber
	&C_2(t)=&\frac{k_{in}k^{trans}}{v_e}\times 
	\\*
	&&\times\left[v(e^{k_{el}t'}-1)e^{-k_{el}t}-u(e^{k_{ep}t'}-1)e^{-k_{ep}t}\right]
\end{eqnarray}
where coefficients $u$ and $v$ are given by,
\begin{eqnarray}\label{eq:eq5}
\nonumber
	&u=&\left[k_{ep}(k_{ep}-k_{el})\right]^{-1}\\
	&v=&\left[k_{el}(k_{ep}-k_{el})\right]^{-1}
\end{eqnarray}

In the solution set, equation \ref{eq:eq4}, the time parameter $t'$  defines the different phases for the progression of the contrast agent. During the wash-in phase, $0\leq t \leq\tau$, where $\tau$ is the contrast infusion time, $t'$ is taken equal to $t$. In the wash-out phase, $t>\tau$, $t'$ is set equal to $\tau$.

\subsection{Fast and Slow Exchange Compartments}

The Brix-Tofts model represents a simple approach to a real system. Several authors \cite {Ambi16,Chen11,Chen111,Wang06} have proposed multi-compartmental models to describe the kinetics of the contrast in the tissue. Among these models, the simplest one corresponds to the assumption that the system can be described by a three-compartment model \cite{Chen11,Wang06} with the following compartmental equations,
\begin{eqnarray}
\label{eq:eq5-1}
\nonumber
	&\frac{dC_{p}}{dt} = &\frac{k_{in}}{v_{p}} - (k_{s}^{trans} + k_{f}^{trans} + k_{el})C_{p} +
	\\
\nonumber
	 &&+\frac{k_{eps}v_{es}}{v_{p}}C_{s} + \frac{k_{epf}v_{ef}}{v_{p}}C_{f}\\
\nonumber
	&\frac{dC_{s}}{dt} = &\frac{k_{s}^{trans}v_{p}}{v_{es}}C_{p} - k_{eps}C_{s}\\
	&\frac{dC_{f}}{dt} = &\frac{k_{f}^{trans}v_{p}}{v_{ef}}C_{p} - k_{epf}C_{f}
\end{eqnarray}

Under the same assumptions made for the Brix-Tofts model, equations (\ref{eq:eq5-1}) have the solution,
\begin{align}
\label{eq:eq5-2}
\nonumber
	&C_{p}(t)=\frac{k_{in}}{v_{p}k_{el}}(e^{k_{el}t'}-1)e^{-k_{el}t}\\
\nonumber
	&C_{s}(t)=\frac{k_{in}k_{s}^{trans}}{v_{es}}\times\\
\nonumber
	&\qquad\quad\times\left[v_{s}(e^{k_{el}t'}-1)e^{-k_{el}t}-u_{s}(e^{k_{eps}t'}-1)e^{-k_{eps}t}\right]\\
\nonumber
	&C_{f}(t)=\frac{k_{in}k_{f}^{trans}}{v_{ef}}\times\\
	&\qquad\quad\times\left[v_{f}(e^{k_{el}t'}-1)e^{-k_{el}t}-u_{f}(e^{k_{epf}t'}-1)e^{-k_{epf}t}\right]
\end{align}
with,
\begin{eqnarray}
\nonumber
	&u_{s,f}=&\left[k_{ep;s,f}(k_{ep;s,f}-k_{el})\right]^{-1}\\
	&v_{s,f}=&\left[k_{el}(k_{ep;s,f}-k_{el})\right]^{-1}
\end{eqnarray}
and similarly, in equations (\ref{eq:eq5-2}) during the wash-in phase, $0\leq t \leq\tau$, where $\tau$ is the contrast infusion time, $t'$ is taken equal to $t$, and in the wash-out phase, $t>\tau$, $t'$ is set equal to $\tau$. The contrast concentration in the tissue is the sum of $C_{s}$ and $C_{f}$.

\subsection{Image acquisition and quantitative analysis} 
$T_{1}$-weighted MRI intensity is related to the extravascular and extracellular contrast concentration $C_{2}$ through the following relation [5],
\begin{equation}\label{eq:eq6}
	S_t\approx S_0(1+FC_2(t))
\end{equation}
which holds if certain conditions apply for the MRI sequence parameters, i.e., $TR\alpha C_2\ll1$ and $TE\beta C_2$, where $\alpha$ and $\beta$ determine the enhancement of longitudinal and transversal relaxation rates, respectively, due to the contrast agent, and $S_0$ is the image intensity without contrast. In general, DCE - MRI protocols are fine-tuned to fulfill the necessary conditions for equation \ref{eq:eq6} to apply, and in such a case, equations for the concentration $C_2$ can be obtained from the image data set as,
\begin{align}
\label{eq:eq7}
\nonumber
	&C_2=A\left[v(1-e^{-k_{el}t})-u(1-e^{-k_{ep}t})\right],  t \leq \tau
	\\
	&C_2=A\left[v(e^{k_{el}\tau}-1)e^{-k_{el}\tau}-u(e^{k_{ep}\tau}-1)e^{-k_{ep}t}\right],  		 	t>\tau
\end{align}

Taking into account that $u$ and $v$ are given by equation \ref{eq:eq5}, only four parameters are needed to fit the model to experimental data: $k_el$, $k_ep$, $\tau$ and $A$. Prostate DCE-MRI was obtained from the Collection Prostate-Diagnosis at The Cancer Imaging Archive (TCIA), National Cancer Institute \cite{Clark13}. 

The image data set was previously registered to the image for the initial condition, i.e., $T_1$-weighted MRI with no contrast, to diminish physiological and involuntary patient movements. 3D rigid image registration was performed using MATLAB Image Processing Toolbox, with Mutual Information \cite{Mattes01} as a similarity metric for the Regular Step Gradient Descent \cite{Nocedal06,vdBom11} optimization algorithm.

\subsubsection{Levenberg-Marquardt fit.}
Quantification of pharmacokinetic parameters was also performed with MATLAB, using least square fitting by the Levenberg-Marquardt algorithm \cite{Leven44,Marq63}. As previously mentioned, only four parameters are needed to fit the model to experimental data; if it is assumed that only the ratio between $u$ and $v$ is used instead of their actual expressions, equation \ref{eq:eq5} and the proportionality coefficient $A$ is incorporated to either $u$ or $v$, then the number of parameters is maintained, and equation \ref{eq:eq7} can be rewritten as,
\begin{align}
\nonumber
	&C_2=A'\left[\frac{k_{ep}}{k_{el}}(1-e^{-k_{el}t})-(1-e^{-k_{ep}t})\right],  	t\leq\tau
	\\
	&C_2=A'\left[\frac{k_{ep}}{k_{el}}(e^{k_{el}\tau}-1)e^{k_{el}t}-(e^{k_{ep}\tau}-1)e^{-k_{ep}t}\right],  	t>\tau
\end{align}

On the other hand, if the two-exponential dependence of the model is preserved, i.e., it is a two-compartment model, some freedom can be gained for the fitting of experimental data and only five parameters are needed,
\begin{align}
\label{eq:eq9}
\nonumber
	&C_2=A_1(1-e^{-k_{el}t})-A_2(1-e^{-k_{ep}t}),  t\leq\tau
	\\
	&C_2=A_1(e^{k_{el}\tau}-1)e^{-k_{el}t}-A_2(e^{k_{ep}\tau}-1)e^{-k_{ep}t},  t>\tau
\end{align}

An example of the application of the model with a different number of fitting parameters is shown in Figure 2. As expected, the model with five parameters, yields the best fit, while the four-parameter model imposes too many restrictions between the parameters avoiding data fitting. Nevertheless, the four-parameter model could be used to confirm the Brix-Tofts model's exact validity within the data noise limits.
\begin{figure}[H]
    \centering
    \includegraphics[width=12cm]{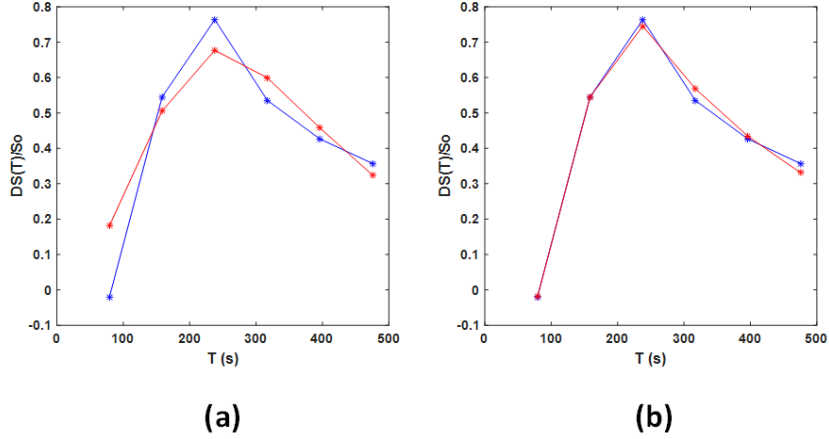}
    \caption{Different data fitting results for the same pharmacokinetic model, fitted curve is indicated in red. (a) Four parameters, (b) Five parameters. In both cases, the Levenberg Marquardt algorithm was used}
    \label{fig:f2}
\end{figure}

\subsubsection{Modified de Prony method fit. Two Compartments Model}
A modified version of the de Prony method \cite{dProny1795} was implemented for multi-echo $T_{2}$-weighted MRI \cite{Mml07} to quantify transversal rate distributions in solid brain tumors. Essentially, the method assumes that the signal or image intensity is described by a superposition of a finite number of decaying exponential functions and it is sampled at regularly fixed time intervals as is the case for DCE-MRI. If the Brix-Tofts model is assumed and following equation \ref{eq:eq9}, voxel intensity is given by,

\begin{align}
\label{eq:eq10}
\nonumber
	p_i=(A_1-A_2)-A_1X_{1}^{i}+A_2X_{2}^{i},  1\leq i\leq n, i\delta t\leq\tau
	\\
	p_{n+j}=B_1X_{1}^{n+j}-B_2X_{2}^{n+j},  1\leq j\leq m, (n+j)\delta t>\tau
\end{align}
where the following definitions apply,
\begin{eqnarray}\label{eq:eq11}
\nonumber 
	&X_1&=e^{-k_{el}\delta t}\\
\nonumber
	&X_2&=e^{-k_{ep}\delta t}
	\\
\nonumber
	&A_1&=F \frac{k_{in}k^{trans}}{v_e} \left[k_{el}(k_{ep}-k_{el})\right]^{-1}
	\\
\nonumber
	&A_2&=F \frac{k_{in}k^{trans}}{v_e} \left[k_{ep}(k_{ep}-k_{el})\right]^{-1}
	\\
\nonumber
	&B_1&=(e^{k_{el}\tau}-1)A_1
	\\
	&B_2&=(e^{k_{ep}\tau}-1)A_2
\end{eqnarray}

As it can be seen in the definitions shown in equation \ref{eq:eq11}, the parameters are not independent if the Brix-Tofts model strictly applies and only a set of four independent parameters is left as discussed in the previous section. In particular, there is some relationship between the parameters,
\begin{equation}\label{eq:eq11a}
	\frac{A_{1}}{A_{2}} = \frac{k_{ep}}{k_{el}}
\end{equation}

Because the shortest sampling interval in DCE-MRI depends on the volume acquisition time, the number of points and the precise definition of $\tau$, which is required for the application of this method, is limited. Nevertheless, as a first approach, each section of the voxel intensity evolution, i.e., wash-in or wash-out, can be analyzed separately and the resulting exponents compared. Let us consider the case of the wash-in section, which according to equation \ref{eq:eq10} contains a time-independent term and can be eliminated by taking differences between consecutive points, $q_i=p_i-p_{i+1}$, for $1\leq i\leq n-1$. The original system, equation \ref{eq:eq10} is transformed to \cite{Mml07},
\begin{equation}
	\left[ \begin{array}{c} q_3 \\ q_4 \end{array} \right]=\left[ \begin{array}{cc} q_2 & -q_1 \\ q_3 & -q_2 \end{array} \right] \left[ \begin{array}{c} X_1+X_2 \\ X_1X_2 \end{array} \right]
\end{equation}
with solutions $Z_1^*\equiv X_1+X_2$, $Z_2^*\equiv X_1 X_2$. The solution for $X_1$ and $X_2$ are obtained by finding the roots of a second-degree polynomial,
\begin{equation}
	X^2-Z_1^*X+Z_2^* =0
\end{equation}
with the condition that the roots must be real and $0<X<1$. Exponents and coefficients are determined straightforwardly by substitution in equations \ref{eq:eq10} and \ref{eq:eq11}. An analogous procedure could be performed for the wash-out section and the resulting exponents and coefficients compared to those determined for the wash-in section. In principle, parameters extracted from each trend, i.e., wash-in and wash-out, should be equal if a strict Brix-Tofts model applies. This procedure is somewhat cumbersome and could be affected by the SNR in the image intensity and patient movement so the matching of parameters seems to be unlikely. The method is then limited only to the analysis of the wash-in or wash-out data \cite{Aponte16}. 

A possible solution to this fact is to use the whole set of data points, assuming a common $\tau$, which can be modeled as the following discretized set of equations,
\begin{align}
\label{eq:eq15-1}
\nonumber
	&p_i = A_1(1-X_{1}^{i})-A_2(1-X_{2}^{i}),  1\leq i\leq n, i\delta t\leq\tau
	\\
\nonumber
	&p_{n+j}=A_1(X_{1}^{j}-X_{1}^{n+j})-\\
	&\qquad\qquad-A_2(X_{2}^{j}-X_{2}^{n+j}),  1\leq j\leq m, (n+j)\delta t>\tau
\end{align}
which can be written in matrix form as
\begin{equation}\label{eq:eq15}
	\left( \begin{array}{c} p_1 \\ \vdots \\ p_i \\ \vdots \\ p_n \\ p_{n+1} \\ \vdots \\ p_{n+j} \\ \vdots \\ p_{n+m} \end{array} \right)= \left( \begin{array}{cc} (1-X_{1}^{1}) & -(1-X_{2}^{1}) \\ \vdots & \vdots \\ (1-X_{1}^{i}) & -(1-X_{2}^{i}) \\ \vdots & \vdots \\ (1-X_{1}^{n}) & -(1-X_{2}^{n}) \\ (X_{1}^{1}-X_{1}^{n+1}) & -(X_{2}^{1}-X_{2}^{n+1}) \\ \vdots & \vdots\\ (X_{1}^{j}-X_{1}^{n+j}) & -(X_{2}^{j}-X_{2}^{n+j}) \\ \vdots & \vdots \\ (X_{1}^{m}-X_{1}^{n+m}) & -(X_{2}^{m}-X_{2}^{n+m}) \end{array} \right) \left( \begin{array}{c} A_1 \\ A_2 \end{array} \right)
\end{equation}
with $X_1>X_2$.

\begin{figure}[!h]
    \centering
    \includegraphics[width=10cm]{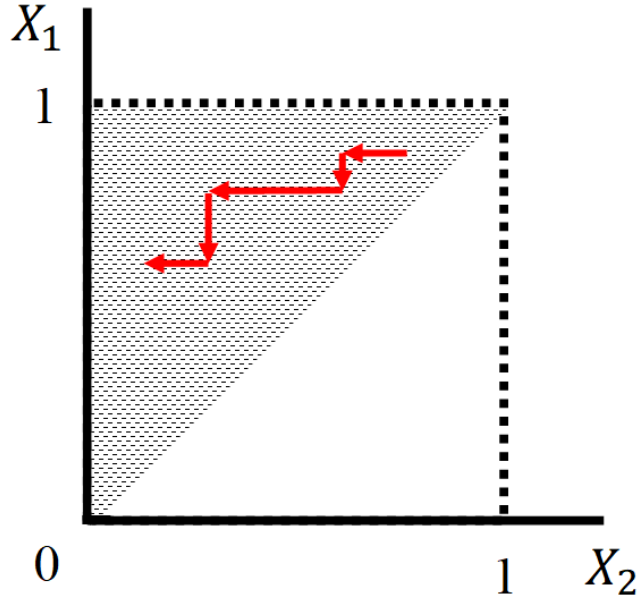}
    \caption{Location of exponential parameters for the set given by equation \ref{eq:eq15}. Dashed areas represent possible solutions. In red, the iterative path followed by a steepest-descent method to obtain an optimal solution for $A_1$, $A_2$ of equation \ref{eq:eq15}.}
    \label{fig:f3}
\end{figure}

Equation \ref{eq:eq15} resembles a modified version of Vandermonde matrix in terms of geometric progressions and can be solved by a nonnegative least-square method for the coefficients $A_1$ and $A_2$, if a pair $X_1$, $X_2$ is given. It is necessary to search the space $(X_1, X_2)$ for an optimal solution of equation \ref{eq:eq15}. To do so these values are selected initially at random within the square of side 1 and subjected to the condition $X_1>X_2$, as shown in Figure \ref{fig:f3}. The best solution for $A_1$, $A_2$, $X_1$, and $X_2$ is obtained by following the steepest-descent optimization method to minimize the residuals of equation \ref{eq:eq15}, A possible path is shown in Figure \ref{fig:f3}.

One of the advantages of the Vandermonde matrix formulation is that it allows for an irregular sampling of the kinetics data as compared to the modified de Prony formulation which requires regular sampling. This fact allows for the implementation of DCE-MRI protocols suited to the particular requirements of pharmacokinetics. 

\subsubsection{Modified de Prony method fit. Three Compartments Model}
\label{sec:sec_3c}
In the case of a three compartments model, the total concentration in the tissue is
\begin{equation}
	C \equiv C_{s} + C_{f}
\end{equation} 
so equations (\ref{eq:eq9}) have to be modified to 
\begin{eqnarray}
\nonumber
	&C  = &A_1(1-e^{-k_{el}t})-\\
\nonumber
	&&\qquad -A_2(1-e^{-k_{eps}t})-A_3(1-e^{-k_{epf}t}),  t\leq\tau \\
\nonumber
	&C  = &A_1(e^{k_{el}\tau}-1)e^{-k_{el}t}-\\
\nonumber	
	&&\qquad \qquad-A_2(e^{k_{eps}\tau}-1)e^{-k_{eps}t}-\\	
	&&\qquad \qquad \qquad \qquad-A_3(e^{k_{epf}\tau}-1)e^{-k_{epf}t}, t>\tau
\end{eqnarray}
and the straightforward modifications to equations (\ref{eq:eq15-1}) and (\ref{eq:eq15}) with the condition $X_{1} > X_{2} > X_{3}$, which in matrix form is,
\begin{equation}
\label{eq:eq16}
	\left( \begin{array}{c} p_1 \\ \vdots \\ p_i \\ \vdots \\ p_n \\ p_{n+1} \\ \vdots \\ p_{n+j} \\ \vdots \\ p_{n+m} \end{array} \right)= \left( \begin{array}{ccc} (1-X_{1}^{1}) & -(1-X_{2}^{1} & -(1-X_{3}^{1}) \\ \vdots & \vdots \\ (1-X_{1}^{i}) & -(1-X_{2}^{i}) & -(1-X_{3}^{i}) \\ \vdots & \vdots \\ (1-X_{1}^{n}) & -(1-X_{2}^{n}) & -(1-X_{3}^{n})\\ (X_{1}^{1}-X_{1}^{n+1}) & -(X_{2}^{1}-X_{2}^{n+1}) & -(X_{3}^{1}-X_{3}^{n+1}) \\ \vdots & \vdots\\ (X_{1}^{j}-X_{1}^{n+j}) & -(X_{2}^{j}-X_{2}^{n+j}) & -(X_{3}^{j}-X_{3}^{n+j}) \\ \vdots & \vdots \\ (X_{1}^{m}-X_{1}^{n+m}) & -(X_{2}^{m}-X_{2}^{n+m}) & -(X_{3}^{m}-X_{3}^{n+m}) \end{array} \right) \left( \begin{array}{c} A_1 \\ A_2 \\ A_3 \end{array} \right)
\end{equation}

\subsection{Tissue classification by semi-qualitative methods}
\begin{figure}[!h]
    \centering
    \includegraphics[width=12cm]{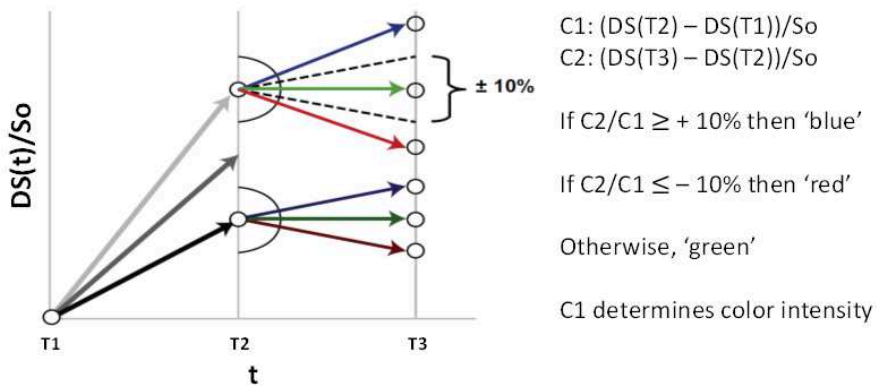}
    \caption{Three-time points method for tissue classification in DCE-MRI \cite{8,9}.}
    \label{fig:f4}
\end{figure}

Tissue classification has been accomplished in DCE-MRI data in different ways \cite{Degani97,Hauth06,Lenkinski11,Fluckiger12,Lavini13,Furman14}, exploiting properties of the data time evolution, and has been extensively used clinically. Among them, the so-called three time point method \cite{Degani97,Hauth06,Fluckiger12,Furman14,Sansone15} has an extended use. The method consists of selecting three temporal points where the time evolution of the contrast agent uptake is represented appropriately. These points are commonly selected as the time for the initial image set, corresponding to an absence of contrast, an intermediate time point, associated with the infusion time $\tau$, and a third point at the end of the contrast washout region. Image intensity differences are calculated for each voxel and are used for tissue classification according to the scheme shown in Figure \ref{fig:f4}.

Color intensity is determined by the image difference $C_{1}$ also shown in Figure \ref{fig:f4}. The selection of the color scheme is made to correlate with histological measurements \cite{9} assuming that `red' means pathological tissue with a Type III kinetics, `blue' means not affected tissue with a Type I kinetics, and `green' applies to those tissues with uncertain conditions, with a Type II kinetics. If this color code scheme is applied to the Brix-Tofts pharmacokinetic model it is possible to obtain a color map for the pharmacokinetic parameters. In the present work, semi-qualitative analysis is used to classify quantitative results.

\section{Results and Discussion}
\subsection{Semi-qualitative results}
\begin{figure}[!h]
    \centering
    \includegraphics[width=12cm]{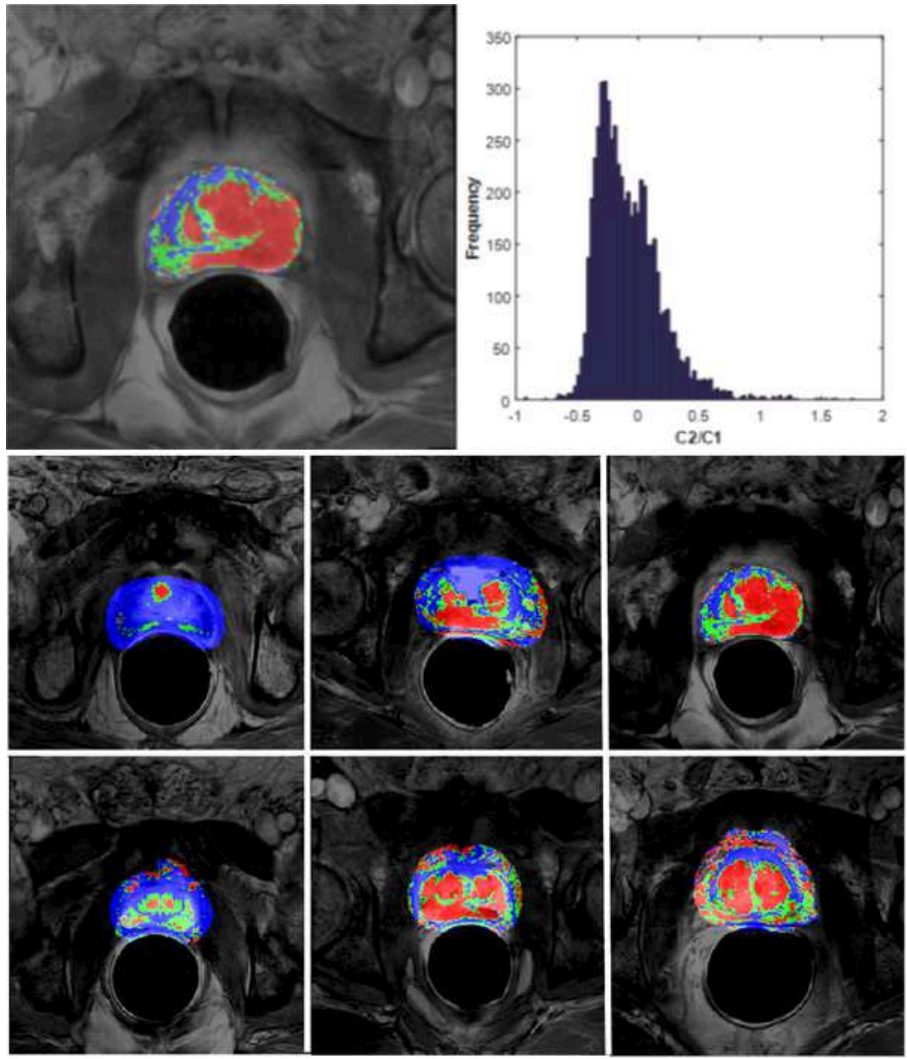}
    \caption{Application of the three-time point method. The top row, left side, superposition of the classified image, and the original one, right side, is the distribution of points according to the $C_2/C_1$ ratio. The bottom rows are examples of the qualitative classification made by the three-time point method.}
    \label{fig:f5}
\end{figure}

Results obtained with the application of the three-time point method are partially shown in Figure \ref{fig:f5}. The prostate was manually segmented and only points within the ROI were evaluated. There was a gradual color transition for all the cases, i.e., red to green to blue, from the tumor lesion's interior to its periphery, as shown in some examples at the bottom rows in Figure \ref{fig:f5}. Spurious points are neglected based on their intensity, i.e., $C_{i}$ value, and local neighborhood.

Some semi-quantitative information can be extracted from the qualitative analysis. The semi-quantitative parameters $C_1$ and $C_2$, defined in Figure \ref{fig:f4} can be used to determine semi-quantitative evaluations that help to characterize the pathology or changes during a therapy follow-up \cite{Wu15,Xiao21,Zeng21,Zhu22}. An example of this approach is shown in the right-hand side of Figure \ref{fig:f5}, where a distribution is obtained for the $C_2⁄C_1$ ratio.

\subsection{Quantitative results}
\subsubsection{Levenberg-Marquardt results.}
\begin{figure}[!h]
    \centering
    \includegraphics[width=12cm]{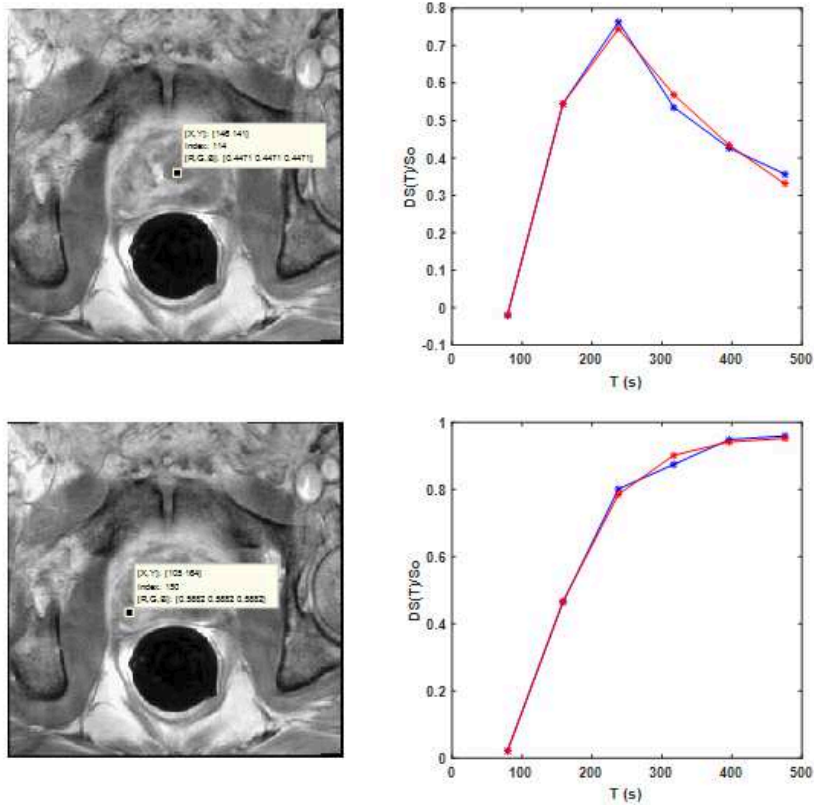}
    \caption{Results of the Levenberg-Marquardt five parameters fitting at two different points: top, a point located within the cancerous lesion, a red region in Figure \ref{fig:f5} with fitted parameters $k_{ep}=0.1321 s^{-1}$ and $k_{el}=0.0034 s^{-1}$; bottom, located in a region outside the cancerous lesion, blue-green (Figure \ref{fig:f5}) with parameters, $k_{ep}=0.0123 s^{-1}$, and $k_{el}=0.00009 s^{-1}$}
    \label{fig:f6}
\end{figure}

The quantitative analysis of the image data set was performed in a previous work \cite{Ruiz17,Aponte16} using only a five-parameter fit of the Brix-Tofts model to the measured data since as can be seen in Figure \ref{fig:f2}, it was substantially better than the four-parameter implementation. Figure \ref{fig:f6} it is shown an example of the results for different points within the prostate, demonstrating the discrimination capacity of the method to establish significant differences between the pharmacokinetic parameters $k_{el}$ and $k_{ep}$, on a voxel by voxel basis.

\begin{figure}[H]
    \centering
    \includegraphics[width=10cm]{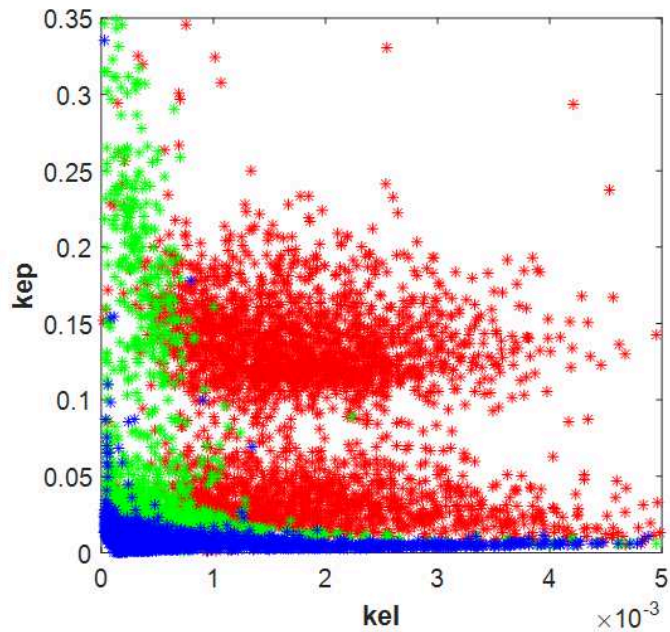}
    \caption{Parameter distributions obtained from Levenberg-Marquardt fit}
    \label{fig:f7}
\end{figure}

The procedure was extended to determine the distribution of these parameters over a region of interest like the one represented in Figure \ref{fig:f5}, as shown in Figure \ref{fig:f7}.

Figure \ref{fig:f7} reflects an interesting result that possibly requires further analysis. It is the apparent difference in point clustering, particularly for those points associated with tumor lesions (red) and undefined and possibly infiltrated tissue (green). This difference can be used to further classify the tumor lesions or their evolution during therapy follow-up. Even though the Levenberg-Marquardt fit is a very reliable method it depends strongly on the actual parameter structure of the pharmacokinetic model as previously discussed and shown in Figure \ref{fig:f2}, making it unsuitable for models that include more than two compartments. In the following, we will limit our discussion to a simpler approach based on the modified de Prony method.

\subsubsection{Modified de Prony method results. Two Compartments Model}
\begin{figure}[!h]
    \centering
    \includegraphics[width=\columnwidth]{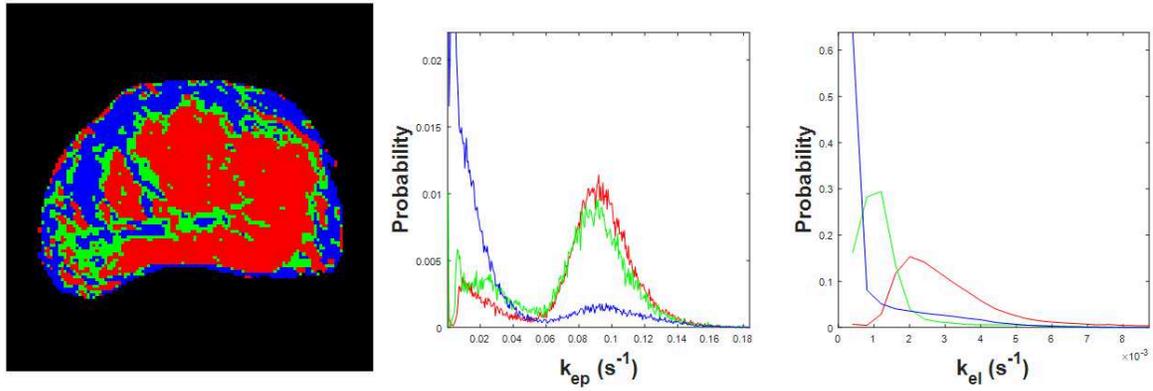}
    \caption{Two compartment model distributions $(k_{ep},k_{el})$. Left, color map; Middle, $k_{ep}$ distribution; Right, $k_{el}$ distribution. Colors are given according to the 3-time-point classification}
    \label{fig:f8}
\end{figure}

\begin{figure}[H]
    \centering
    \includegraphics[width=12cm]{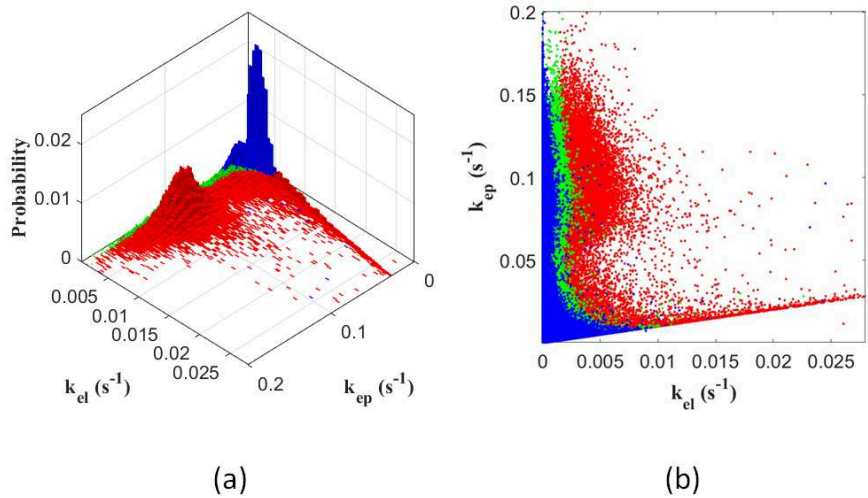}
    \caption{Distributions $(k_{ep},k_{el})$. (a) 2D histograms; (b) $(k_{ep},k_{el})$ plot. Colors are given according to the 3-time-point classification}
    \label{fig:f9}
\end{figure}

The analysis of the quantitative results based on a Levenberg-Marquardt indicates the validity of the Brix-Tofts two-compartment model in the sense that it applies to a specific voxel. When the whole set of voxels is analyzed, as shown in Figure \ref{fig:f7}, it is evident that there are two distinct regions for the kinetic parameters that can be understood as the existence of fast exchange compartments ($k_{ep}$ high) and slow compartments ($k_{ep}$ low), allowing for the proposition of a three-compartment model as previously discussed. The application of the two-compartment model using the modified de Prony algorithm proposed in this work reveals the same behavior as shown by the results in Figure \ref{fig:f8} and Figure \ref{fig:f9}

\begin{figure}[H]
    \centering
    \includegraphics[width=12cm]{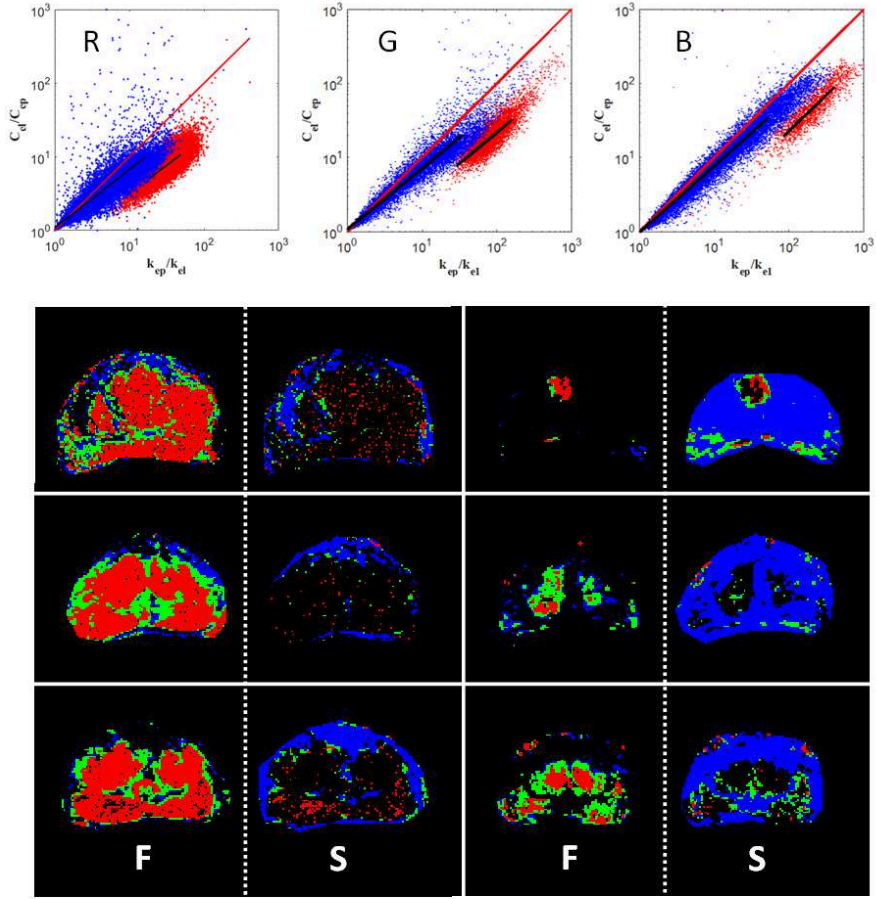}
    \caption{Top: Data trends for $\frac{C_{el}}{C_{ep}}$ (or $\frac{A_1}{A_2}$, according to equation \ref{eq:eq11}) and $\frac{k_{ep}}{k_{el}}$ showing how these ratios are correlated. Red straight lines represent the condition $\frac{C_{el}}{C_{ep}}=\frac{k_{ep}}{k_{el}}$ which corresponds to a pure two-compartment model. Fast exchange compartments (red dots) are differentiated from slow exchange compartments (blue dots). Black lines identify trends in either the fast or slow compartments. Bottom rows: Examples of tissue classification according to the type of exchange compartments. $\textbf{(F)}$ Fast exchange compartments ($k_{ep}$ high). $\textbf{(S)}$ Slow exchange compartments ($k_{ep}$ low). Colors are given according to the 3-time-point classification. Notice that the majority of the fast exchange component is associated with tumor activity, while the slow exchange component is associated with benign tissue}
    \label{fig:f10}
\end{figure}

In particular, in Figure \ref{fig:f8}, the distribution for the $k_{ep}$ parameter is bi-modal with a high exchange region mostly populated by voxels associated with the presence of tumor (red and green colors) while the slow exchange region is predominantly composed of voxels that belong to benign tissue (blue color). This feature can be greatly emphasized if the logarithm of the quotient $\frac{A_1}{A_2}$ is plotted against the logarithm of $\frac{k_{ep}}{k_{el}}$, as shown in Figure \ref{fig:f10}. 

\begin{figure}[H]
    \centering
    \includegraphics[width=12cm]{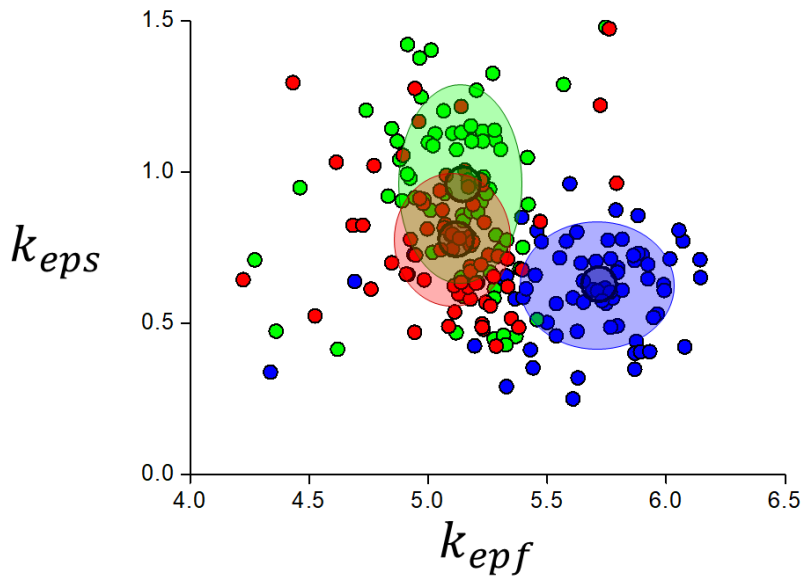}
    \caption{Data points in $(k_{epf},k_{eps})$ space. Big circles represent average values from Table \ref{tab:tab1}. Color-shaded areas represent points within a standard deviation. Colors are given according to the 3-time-point classification}
    \label{fig:f10_1}
\end{figure}

If only one type of compartment is available for the tissue, corresponding to a pure Brix-Tofts model, the plot should behave as a straight line derived from equation \ref{eq:eq11a} and shown as a red line in Figure \ref{fig:f10}. The real trend of the data supports what is observed in the $k_{ep}$ distribution shown in Figure \ref{fig:f8} with evident bi-modal behavior. All these observations suggest that the description of the system should be addressed with a three or more-compartment model. The spatial distribution of the fast and slow exchange compartments is shown at the bottom of Figure \ref{fig:f10}. Specific values of the kinetic parameters for the different kinetic types as classified by the three-time point method are summarized in Table \ref{tab:tab1} and Figure \ref{fig:f10_1}.

\begin{table}[!h]
\caption{Fast and Slow Exchange Kinetic Parameters}
\label{tab:tab1}
\centering
\renewcommand{\arraystretch}{1.5}
\setlength{\tabcolsep}{5pt}
\begin{tabular}{|l|l|l|l|}
\hline
Classification &
$k_{epf} (min^{-1})$ & $k_{eps} (min^{-1})$ & $k_{el} (min^{-1})$ \\
\hline
Type I & $5.72 \pm 0.32$ & $0.63 \pm 0.17$ & $0.04 \pm 0.05$ \\
Type II & $5.15 \pm 0.25$ & $0.96 \pm 0.31$ & $0.05 \pm 0.05$ \\
Type III & $5.12 \pm 0.24$ & $0.78 \pm 0.22$ & $0.12 \pm 0.08$ \\
\hline
\multicolumn{4}{p{280pt}}{Average values for the kinetic parameters in the regions classified by the three-time point method. Type I corresponds to benign tissue and Type III to definitely malignant tissue. Type II corresponds to an intermediate condition, possibly related to tumor invasion.}\\
\end{tabular}
\end{table}

It is important to remark that fast exchange compartments are mostly associated with Type II and Type III kinetics while Type I kinetics corresponds to slow exchange compartments. This allows for an effective $\langle k_{ep} \rangle$ defined as a weighted average between both compartment types. The result is summarized in Table \ref{tab:tab2}.

\begin{table}[!h]
\caption{Proportion of fast exchange compartments and average $\langle k_{ep} \rangle$}
\label{tab:tab2}
\centering
\renewcommand{\arraystretch}{1.5}
\setlength{\tabcolsep}{5pt}
\begin{tabular}{|l|l|l|}
\hline
Classification & $f_{fast}$ &  $\langle k_{ep} \rangle (min^{-1})$\\
\hline
Type I & $0.20 \pm 0.10$ & $1.75 \pm 0.60$ \\
Type II & $0.70 \pm 0.22$ & $4.00 \pm 0.99$ \\
Type III & $0.77 \pm 0.20$ & $4.21 \pm 0.97$ \\
\hline
\end{tabular}
\end{table}

\begin{figure}[!h]
    \centering
    \includegraphics[width=12cm]{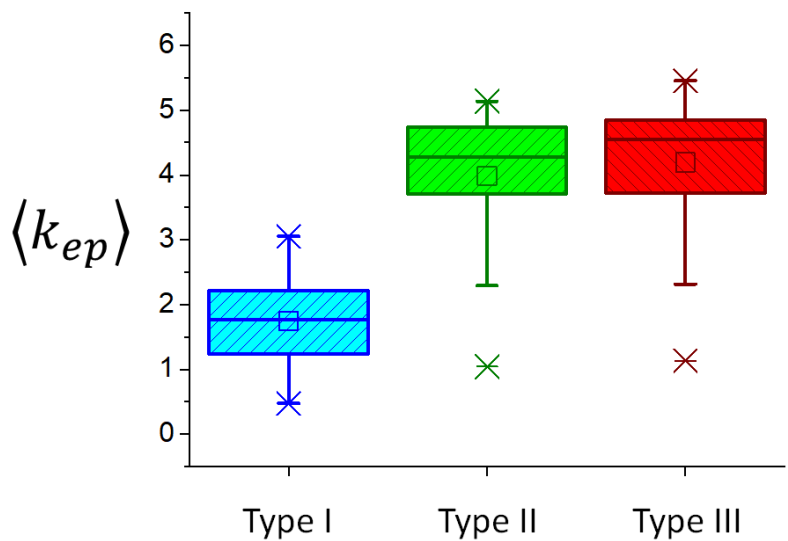}
    \caption{Average value $k_{ep}$ for different kinetic types. Colors are given according to the 3-time-point classification}
    \label{fig:f11_0}
\end{figure}

\subsubsection{Modified de Prony method results. Three Compartments Model}
\begin{figure}[!h]
    \centering
    \includegraphics[width=\columnwidth]{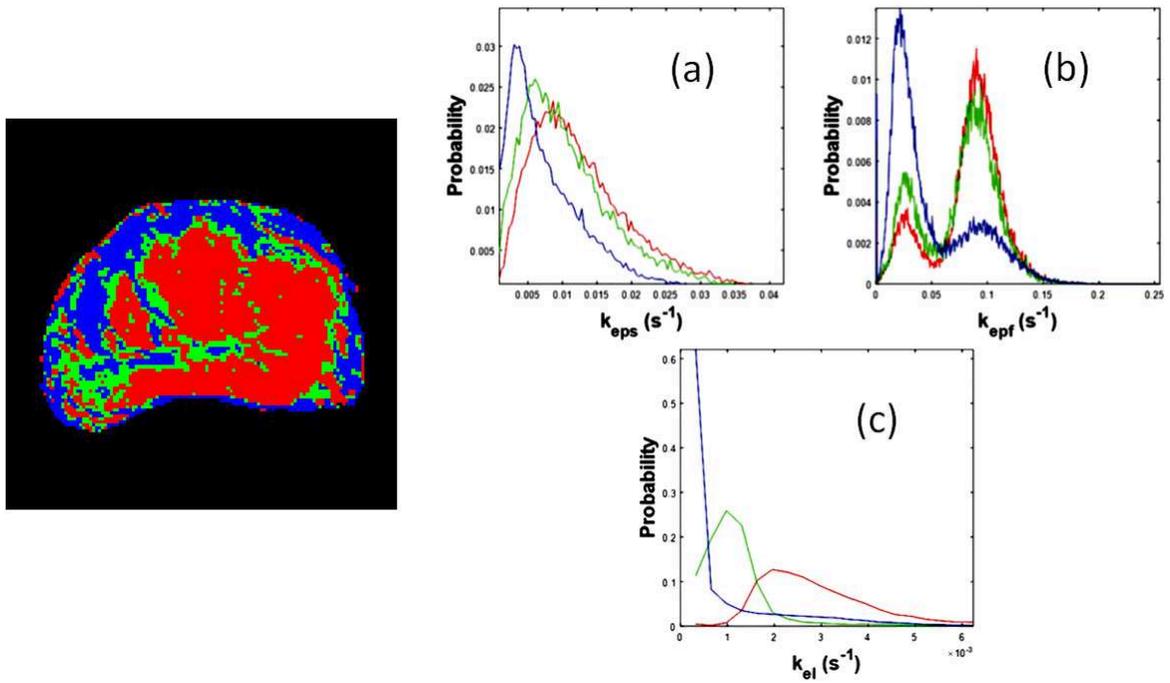}
    \caption{Distributions of the kinetic parameters using a three compartments model. (a) $k_{eps}$ for slow exchange,(b) $k_{epf}$ for fast exchange and (c) $k_{el}$. Colors are given according to the 3-time-point classification}
    \label{fig:f12}
\end{figure}

\begin{figure}[!h]
    \centering
    \includegraphics[width=12cm]{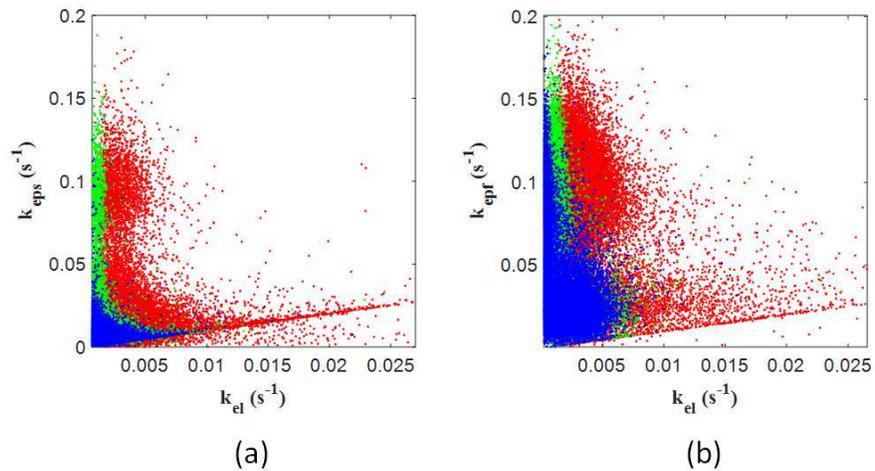}
    \caption{Distributions of the kinetic parameters using a three compartments model in parameter space. (a) $(k_{eps},k_{el})$ for slow exchange and (b) $(k_{epf},k_{el})$ for fast exchange. Colors are given according to the 3-time-point classification}
    \label{fig:f13}
\end{figure}

The previous section discussed the analysis with a pure Brix-Tofts model (two compartments system) with a clear indication of the presence of fast and slow exchange compartments. With the assumption of a three-compartment model, as stated in section \ref{sec:sec_3c}, a more detailed picture of the kinetic parameters is obtained, as shown in Figure \ref{fig:f12}, with some additional structure for the fast exchange parameter, $k_{epf}$. Also, in Figure \ref{fig:f13} it is shown the distribution of kinetic parameters in the $(k_{el},k_{eps})$ and $(k_{el},k_{epf})$ spaces. Nevertheless, comparing the $k_{epf}$ distribution, Figure \ref{fig:f12}b with the $k_{ep}$ distribution in Figure \ref{fig:f8}, they are both bi-modal distributions and very similar indeed, so it can be concluded that the addition of an extra set of parameters to the model does not improve the overall picture increasing notoriously computational times.

\section*{Conclusions}
\label{sec:conclusions}

In the present work, it has been shown a general methodology to classify and quantify tumor lesions in the prostate. The classification of the tissues was performed successfully by the implementation of a suitable three-time point method algorithm. This classification was used as a reference for the analysis of quantitative results. The quantitative analysis was successfully performed by a novel point-wise method based on the two-compartment Brix-Tofts model and validated with the Levenberg-Marquardt optimization method. The results confirmed the existence of fast exchange compartments associated with tumor activity and slow exchange compartments associated with benign tissue. In any case, the combined qualitative and quantitative analysis can be used to establish differences for a patient undergoing therapy or pathology progression. In particular, some results suggest the future possibility of combining the qualitative and quantitative analysis into a single representation for prostate cancer evaluation and therapy follow-up.

\section*{Acknowledgment}
The authors would like to thank the National Insitute for Bioengineering, INABIO, at Universidad Central de Venezuela, Venezuela,  Universidad Nacional Pedro Henríquez Ureña, Dominican Republic and Fundación Arturo López Pérez, Chile, for providing the environment for the realization of this work. Also, we would like to express our gratitude to the scientific community at the Physics and Mathematics in Biomedicine Consortium which provided helpful discussions during the realization of this work.


\bibliography{report}   
\bibliographystyle{spiejour}   


\listoffigures
\listoftables

\end{spacing}
\end{document}